\documentclass[twoside,12pt]{article}
\usepackage [english]{babel}
\usepackage{graphicx}
\usepackage[normalem]{ulem} 
\usepackage{anysize}
\usepackage{fancyhdr}
\usepackage{calc}
\usepackage{geometry}
\usepackage{epsfig}
\usepackage{amssymb}
\usepackage{amsmath}
\usepackage{indentfirst}
\usepackage{chngcntr}

\def\Journal#1#2#3#4{{#1} {#2} (#4) #3 }

\def\NPA{{\em Nucl. Phys.} A}

\def\PLB{{\em Phys. Lett.} B}

\def\PL{{\em Phys. Lett.}}
\def\PRL{\em Phys. Rev. Lett.}

\def\PRC{{\em Phys. Rev.} C}

\newcommand{\be}{\begin{equation}}
\newcommand{\ee}{\end{equation}}
\newcommand{\bea}{\begin{eqnarray}}
\newcommand{\eea}{\end{eqnarray}}

\begin{document}

\title{ \vspace{1cm}Search for He-$\eta$ bound states with the WASA-at-COSY facility}
\author{M.\ Skurzok,$^{1}$ P.\ Moskal,$^{1,2}$ W. Krzemie\'n$^{1}$\\ 
\\
$^1$M. Smoluchowski Institute of Physics, Jagiellonian University, Cracow, Poland\\
$^2$Institut fur Kernphysik, Forschungszentrum J\"ulich, Germany}

\maketitle

\begin{abstract} 
The existence of $\eta$-mesic nuclei in which the $\eta$ meson is bound in a nucleus by means of the strong interaction was postulated already in 1986, albeit not yet confirmed it by experiment. The discovery of this new kind of an exotic nuclear matter would be very important as it might allow for a better understanding of the $\eta$ meson structure and its interaction with nucleons. 
The search for $\eta$-mesic helium ($^{4}\hspace{-0.03cm}\mbox{He}$-$\eta$) is carried out with high statistics and high acceptance with the WASA detector, installed at the cooler synchrotron COSY of the Research Center J\"ulich.
The search is conducted via the measurement of the excitation function for selected decay channels of the $^{4}\hspace{-0.03cm}\mbox{He}$-$\eta$ system. 
In the experiment, performed in November 2010, two reactions $dd\rightarrow(^{4}\hspace{-0.03cm}\mbox{He}$-$\eta)_{bs}\rightarrow$ $^{3}\hspace{-0.03cm}\mbox{He} p \pi{}^{-}$ and $dd\rightarrow(^{4}\hspace{-0.03cm}\mbox{He}$-$\eta)_{bs}\rightarrow$ $^{3}\hspace{-0.03cm}\mbox{He} n \pi{}^{0}$ were measured with a beam momentum ramped from 2.127GeV/c to 2.422GeV/c. The report includes the description of the experimental method and 
status of the measurement.
\end{abstract}

\section{Introduction}

As early as 1985, Bhalerao and Liu~\cite{Bhalerao_Liu} performed a coupled-channel analysis of
the $\pi N \rightarrow \pi N$, $\pi N \rightarrow \pi \pi N$, and $\pi N \rightarrow \eta N$ reactions in the close-to-threshold region,
and discovered that the interaction between the nucleon and the $\eta$ meson is attractive.
This finding inspired Haider and Liu to postulate the existence of $\eta$-mesic nuclei~\cite{HaiderLiu1},
in which the neutral $\eta$ meson might be bound with nucleons via the strong
interaction. The existence of $\eta$-mesic bound states would allow to investigate the interaction of the $\eta$ meson and nucleons inside a nuclear matter.~Moreover, it would provide information about the $\mbox{N}^{*}(1535)$ resonance~\cite{Jido} and about the $\eta$ meson properties in a nuclear matter~\cite{InoueOset}, as well as about contribution of the flavour singlet component of the quark-gluon wave function of the $\eta$ meson~\cite{BassThomas, BassTom}. 
According to theoretical considerations, the formation of the $\eta$-mesic nucleus can only take place if the real part of the $\eta$-nucleus scattering length is negative (attractive nature of the interaction), and the magnitude of the real part is greater than the magnitude of the imaginary part~\cite{HaiderLiu2}:

\begin{equation}
|Re(a_{\eta-nucleus})|>|Im(a_{\eta-nucleus})|.\label{eq:eq1}
\end{equation}\\

\noindent
A wide range of possible values of the s-wave $\eta$N scattering length, from $a_{\eta N}$=(0.27 + 0.22i) fm up to $a_{\eta N}$=(1.05 + 0.27i) fm, calculated for hadronic- and photoproduction of the $\eta$ meson has not excluded the formation of $\eta$-nucleus bound states for a light nuclei as $^{3,4}\hspace{-0.03cm}\mbox{He}$, T~\cite{Wilkin1,WycechGreen} and even for deuteron~\cite{Green}. 
Those bound states have been searched for in many experiments~\cite{Machner,Sokol1,Gillitzer,Berger,Mayer,Mersmann,Smyrski1}, however none of them gave empirical confirmation of their existence. There are only a promising experimental observations which might be interpreted as~\mbox{indications} of $\eta$-mesic nuclei. For example, experimental observations which might suggest the possibility of~the existence of the $^{3}\hspace{-0.03cm}\mbox{He}$-$\eta$ bound system were found by \mbox{SPES-4}~\cite{Berger}, \mbox{SPES-2}~\cite{Mayer}, ANKE~\cite{Mersmann} and \mbox{COSY-11}~\cite{Smyrski1} collaborations. 
The data analysis of the close to threshold SPES-4 and SPES-2 measurements of the total cross \mbox{section} led to the determination of the~$\eta^{3}\hspace{-0.03cm}\mbox{He}$ scattering length which, within the experimental errors, fulfill the condition given by Eq.~\ref{eq:eq1}. In the experiments conducted at the COSY synchrotron the momentum ramping technique of the deuteron beam was used that allows to reduce the systematic uncertainties~\cite{MoskalSmyrski}.~The beam was accelerated slowly and linearly in~time, from excess energy of~\mbox{Q=-5.05 MeV} up~to \mbox{Q=11.33 MeV} in~case~of~the ANKE experiment~\cite{Mersmann}, while during the \mbox{COSY-11} experiment~\cite{Smyrski1} the~momentum was varied in the range corresponding to the excess energy from \mbox{Q=-10 MeV} to \mbox{Q=9 MeV}. Both collaborations performed the measurement of~the~excitation function and differential cross section of~the \mbox{$dp\rightarrow$ $^{3}\hspace{-0.03cm}\mbox{He}$-$\eta$} reaction close to the~kinematical threshold. Data taken by the COSY-11 group were used to search for a signal of a $^{3}\hspace{-0.03cm}\mbox{He}$-$\eta$ bound state below the $\eta$ production threshold of the $dp\rightarrow ppp\pi^{-}$ and $dp\rightarrow$ $^{3}\hspace{-0.03cm}\mbox{He} \pi^{0}$ reactions~\cite{Smyrski3,Krzemien1,Smyrski2}, while the measurements above the threshold enabled the study of the forward-backward asymmetries of the differential cross sections and the extraction of the $\eta$-$^{3}\hspace{-0.03cm}\mbox{He}$ scattering length. The data of both groups~\cite{Mersmann,Smyrski1} show a variation in the phase of the s-wave amplitude in the near-threshold region, consistent with possible existence of a bound state~\cite{Wilkin2}.


The first direct experimental indications of light $\eta$-nucleus bound states were observed in $\eta$ photoproduction $\gamma^{3}$\hspace{-0.03cm}$\mbox{He}\rightarrow \pi^{0}pX$ which was carried out by the TAPS collaboration~\cite{Pfeiffer}. The~measurements of~the excitation functions for two ranges of~the~relative angle between $\pi^{0}$ and proton were carried~out.~It appeared that a~difference between those excitation curves in~the~center-of-mass frame revealed an~enhancement just below the threshold of the $\gamma^{3}\hspace{-0.03cm}\mbox{He}\rightarrow$ $^{3}\hspace{-0.03cm}\mbox{He}$-$\eta$ reaction which was interpreted as a~possible signature of a~{$^{3}\hspace{-0.03cm}\mbox{He}$-$\eta$} bound state where $\eta$ meson captured by one of~nucleons inside helium forms an intermediate $S_{11}(1535)$ resonance which decays into $\pi^{0}$-$p$ pair. However, it was shown later that the result may be an artefact due to the strong influence of the resonances on the shape of the excitation function~\cite{Krusche}. Therefore, a new high-statictics exclusive measurement is necessary.\\

\section{Experiment}
\begin{figure}[h]
\centering
\includegraphics[width=15.5cm,height=7.5cm]{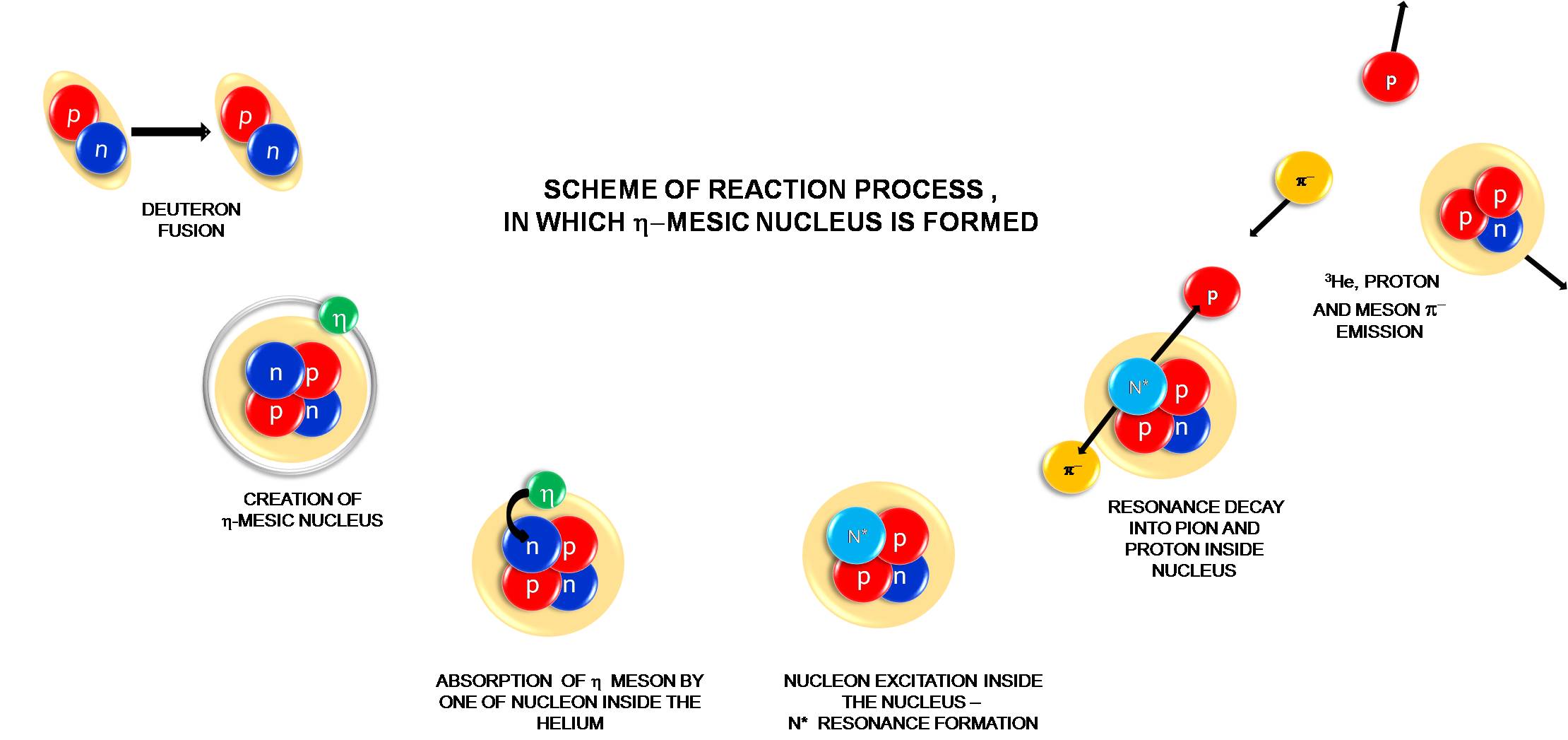}
\caption{Reaction process of the ($^{4}\hspace{-0.03cm}\mbox{He}$-$\eta)_{bs}$ production and decay.\label{fig1}}
\end{figure}

The measurement of the $^{4}\hspace{-0.03cm}\mbox{He}$-$\eta$ bound states is performed with high precision at the COSY accelerator with the WASA detection system.~Signatures of the \mbox{$\eta$-mesic} nuclei are searched for in the excitation function of specific decay channels of the \mbox{$^{4}\hspace{-0.03cm}\mbox{He}$-$\eta$} system, formed in deuteron-deuteron collision~\cite{Moskal1,Krzemien}.~The measurement is performed for beam momenta varying continously around the $\eta$ meson production threshold.~The~beam ramping technique allows to reduce the systematic uncertainities.~The \mbox{existence} of the bound system should manifest itself as a resonance structure in the excitation curve of eg. $dd\rightarrow(^{4}\hspace{-0.03cm}\mbox{He}$-$\eta)_{bs}\rightarrow$ $^{3}\hspace{-0.03cm}\mbox{He} p \pi{}^{-}$ reaction
below the $dd\rightarrow$ $^{4}\hspace{-0.03cm}\mbox{He}$-$\eta$ reaction threshold. 
The kinematics of the reaction is schematically presented in Fig.~\ref{fig1}.~The deuteron beam - deuteron target collision leads to fusion into a $^{4}\hspace{-0.03cm}\mbox{He}$ nucleus in a bound state with the $\eta$ meson. The $\eta$ meson can be absorbed by one of the nucleons inside the helium and may propagate in the nucleus via consecutive excitation of nucleons to the $\mbox{N}^{*}(1525)$ state~\cite{Sokol} until the resonance decays into a pion-proton pair emitted by the nucleus~\cite{Moskal4,KrzeMosSmy}. The relative angle between p and $\pi^{-}$ is equal to
$180^\circ$ in the  $\mbox{N}^{*}$ reference frame and it is smeared by about $30^\circ$ in the center-of-mass frame due to the Fermi motion of the nucleons inside the helium nucleus. 

\begin{figure}[h]
\centering
\includegraphics[width=9.0cm,height=6.0cm]{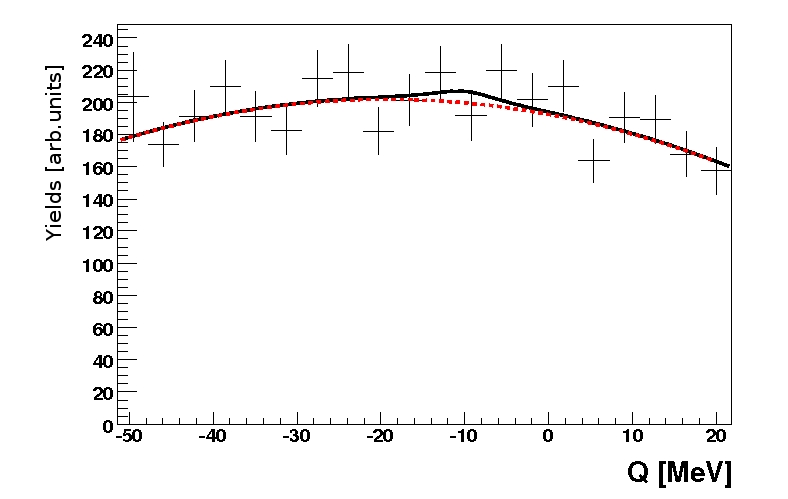}
\caption{Preliminary excitation function for the $dd\rightarrow$ $^{3}\hspace{-0.03cm}\mbox{He} p \pi{}^{-}$ reaction obtained by normalizing events selected in individual excess energy intervals by corresponding integrated luminosities. The solid line represents a fit with second order polynomial combined with a Breit-Wigner function with fixed binding energy and a width equal to -10 and 10 MeV, respectively. The dotted line corresponds to the contribution from the second order polynomial in the performed fit.\label{fig2}}
\end{figure}

In June 2008, the search for the $^{4}\hspace{-0.03cm}\mbox{He}$-$\eta$ bound state was performed by measuring the excitation function of the $dd\rightarrow$ $^{3}\hspace{-0.03cm}\mbox{He} p \pi{}^{-}$ reaction near the $\eta$ production threshold.~During the experimental run the momentum of the deuteron beam was varied continuously within each acceleration cycle from 2.185~GeV/c to 2.400~GeV/c, crossing the kinematic
threshold for the $\eta$ production in the $dd\rightarrow$ $^{4}\hspace{-0.03cm}\mbox{He} \eta$ reaction at~2.336 GeV/c. This range of beam momenta corresponds to a variation of $^{4}\hspace{-0.03cm}\mbox{He}$-$\eta$ excess energy from -51.4 MeV to 22 MeV. 
The excitation function was determined after applying cuts on the p and $\pi^{-}$ kinetic energy distribution and the $p - \pi^{-}$ opening angle in the CM system~\cite{Krzemien_PhD}. The result is shown in Fig.~\ref{fig2}.

\noindent
The relative normalization of the $dd\rightarrow$ $^{3}\hspace{-0.03cm}\mbox{He} p \pi{}^{-}$ excitation function was based on quasi-elastic proton-proton scattering.
The excitation function does not show a structure which could be interpreted as a resonance originating from the decay of an $\eta$-mesic $^{4}\hspace{-0.03cm}\mbox{He}$.

In the experiment, in November 2010, two channels were measured:  $dd\rightarrow(^{4}\hspace{-0.03cm}\mbox{He}$-$\eta)_{bs}\rightarrow$ $^{3}\hspace{-0.03cm}\mbox{He} p \pi{}^{-}$ and  $dd\rightarrow(^{4}\hspace{-0.03cm}\mbox{He}$-$\eta)_{bs}\rightarrow$ $^{3}\hspace{-0.03cm}\mbox{He} n \pi{}^{0} \rightarrow$ $^{3}\hspace{-0.03cm}\mbox{He} n \gamma \gamma$. The measurement was performed with a beam momentum ramping from 2.127GeV/c to 2.422GeV/c, corresponding to the range of the excess energy \mbox{Q$\in$(-70,30)~MeV}. 

\begin{figure}[h!]
\centering
\includegraphics[width=8.0cm,height=7.5cm]{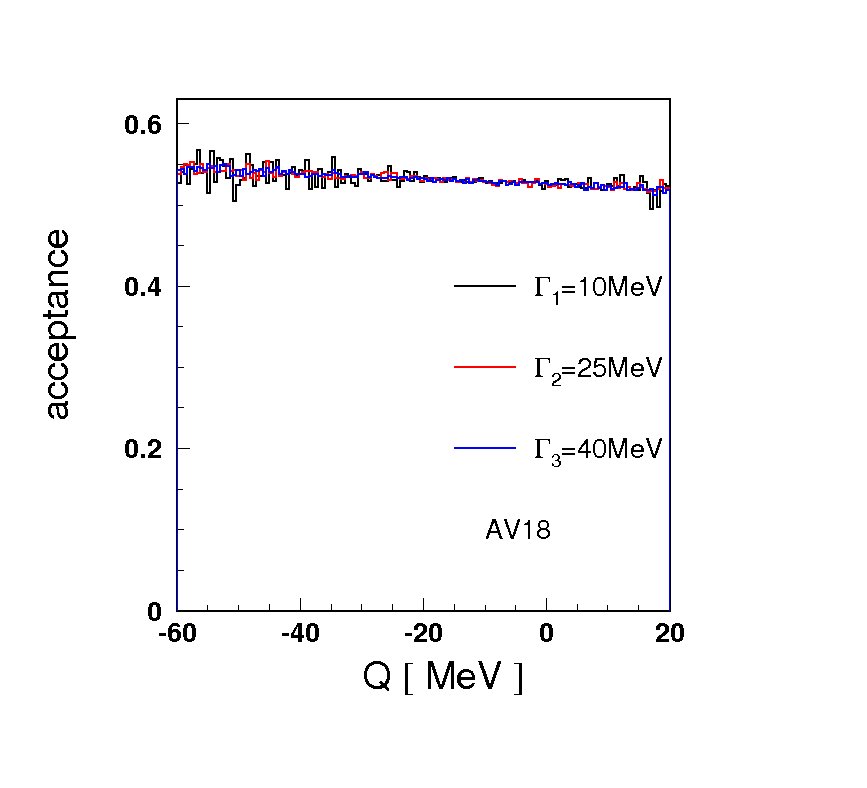} \hspace{-1.7cm} \includegraphics[width=8.0cm,height=7.5cm]{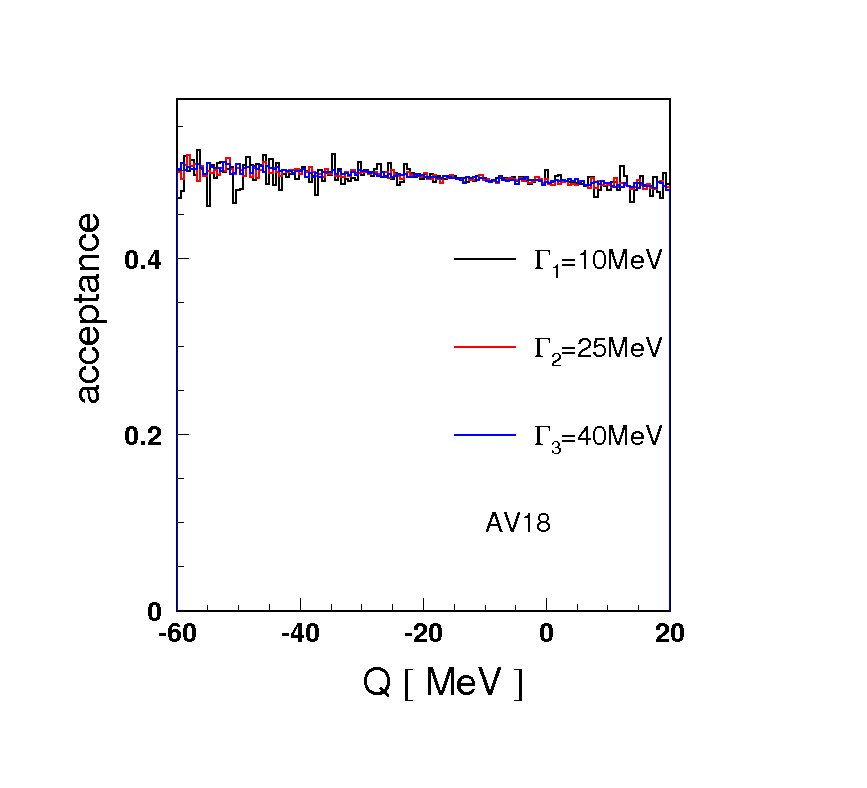}
\vspace{-0.5cm}
\caption{Geometrical acceptances of the WASA-at-COSY detector for the $dd\rightarrow(^{4}\hspace{-0.03cm}\mbox{He}$-$\eta)_{bs}\rightarrow$ $^{3}\hspace{-0.03cm}\mbox{He} p \pi{}^{-}$ (left) and $dd\rightarrow(^{4}\hspace{-0.03cm}\mbox{He}$-$\eta)_{bs}\rightarrow$ $^{3}\hspace{-0.03cm}\mbox{He} n \pi{}^{0} \rightarrow$ $^{3}\hspace{-0.03cm}\mbox{He} n \gamma \gamma$ reaction (right). The acceptance is calculated for different widths of the presumed bound state and using the AV18 potential model description of the nucleon momentum distribution inside $^{4}\hspace{-0.03cm}\mbox{He}$. \label{reakcja_1}}
\end{figure}

\indent For both reactions the geometrical acceptance of the detector as a function of excess energy Q was determined in simulations. The acceptance is presented in Fig.~\ref{reakcja_1} for different widths of the presumed bound state and using the AV18 model description of the nucleon momentum distribution inside the $^{4}\hspace{-0.03cm}He$ nuclei. The detailed description of the simulations is presented in Ref.~\cite{Skurzok}. 

\vspace{0.3cm}

\noindent
The acceptance is almost a constant function of the excess energy and its average value is about 53\% and 50\% for \mbox{$dd\rightarrow(^{4}\hspace{-0.03cm}\mbox{He}$-$\eta)_{bs}\rightarrow$ $^{3}\hspace{-0.03cm}\mbox{He} p \pi{}^{-}$} and $dd\rightarrow(^{4}\hspace{-0.03cm}\mbox{He}$-$\eta)_{bs}\rightarrow$ $^{3}\hspace{-0.03cm}\mbox{He} n \pi{}^{0} \rightarrow$ $^{3}\hspace{-0.03cm}\mbox{He} n \gamma \gamma$, respectively. The high acceptance values allow high statistics measurements of these final states~\cite{Krzemien_PhD}.

\indent During~the~experiment in 2010,~data were effectively taken for about~155 hours. The average luminosity was estimated based on the trigger used for the elastic proton-proton scattering and is about L=8.15$\cdot10^{30} cm^{-2} s^{-1}$. Taking into account the fact that two reactions were measured, in total more than 40 times higher statistics were collected than in the experiment carried out in 2008.

\section{Conclusion}
The search for $^{4}\hspace{-0.03cm}\mbox{He}$-$\eta$ bound states is performed with the WASA-at-COSY
facility by taking advantage of its very high acceptance for the proton-$\pi^{-}$ and neutron-$\pi^{0}$ pairs as well as
very good identification of helium nuclei. In addition, WASA allows an exclusive measurement of the $dd\rightarrow$ $^{3}\hspace{-0.03cm}\mbox{He} p \pi{}^{-}$ reaction.\\
\indent At present the data analysis is in progress. In the optimistic case, the statistics could be sufficient to observe a signal from the $\eta$-mesic
helium and in a pessimistic scenario the upper limit of the cross section for the $^{4}\hspace{-0.03cm}\mbox{He}$-$\eta$ bound state production will be decreased by a factor of about six.

\section{Acknowledgements}

The work is supported by the Foundation for Polish Science - MPD program, co-financed by the European
Union within the European Regional Development Fund, by the Polish National Science Center through grants No. 0320/B/H03/2011/40 and 2011/01/B/ST2/00431, by the European Commission under the 7th Framework Programme through the 'Research Infrastructures' action of the 'Capacities' Programme (FP7-INFRASTRUCTURES-2008-1, Grant Agreement N. 227431) and by the FFE grants of the Research Center J\"ulich.\\

\end{document}